\documentclass[preprint,pra,showpacs,10pt,onecolumn]{revtex4}%
\usepackage{amsfonts}
\usepackage{amsmath}
\usepackage{amssymb}
\usepackage{graphicx}%
\begin{document}
\title{Preparation of cluster states and \textit{W states} with superconducting-
quantum-interference-device qubits in cavity QED}
\author{X.L. Zhang$^{1,2}$}
\email{xili-zhang@hotmail.com}
\author{K.L. Gao$^{1}$}
\email{klgao@wipm.ac.cn}
\author{M. Feng$^{1}$}
\email{mangfeng1968@yahoo.com}
\affiliation{$^{1}$State Key Laboratory of Magnetic Resonance and Atomic and Molecular
Physics, Wuhan Institute of Physics and Mathematics, Chinese Academy of
Sciences, Wuhan 430071, China}
\affiliation{$^{2}$Graduate School of the Chinese Academy of Science, Bejing 100049, China}


\begin{abstract}
We propose schemes to create cluster states and \textit{W states} by many
superconducting-quantum-interference-device (SQUID) qubits in cavities under
the influence of the cavity decay. Our schemes do not require auxiliary
qubits, and the excited levels are only virtually coupled throughout the
scheme, which could much reduce the experimental challenge. We consider the
cavity decay in our model and analytically demonstrate its detrimental
influence on the prepared entangled states.

\end{abstract}
\maketitle

Entanglement is an essential resource for testing quantum nonlocality and also
for quantum information processing. Although many proposals have been put
forward, the achievement of multi-partite entangled states are still
challenging experimentally due to decoherence as well as the limitation of
current techniques. This work focuses on a combinatory system with
superconducting devices embedded in a cavity. Among a variety of qubit
candidates, superconducting devices such as Josephson junction circuits
\cite{1,2}, Josephson junctions \cite{3,4,5}, Cooper pair boxes \cite{6,7},
and superconducting quantum interference devices (SQUIDs) \cite{8,9,10,11,12},
have drawn particular interests due to their potential for integrated devices.
On the other hand, cavities are currently considered as excellent devices to
achieve multiqubit entanglement for atoms, ions, quantum dots, and other
charge qubits \cite{13,14,15,16}.

Our goal is to present ways for creating multi-partite entanglement of many rf
SQUID qubits in cluster states and in \textit{W states }in cavity QED. The
cluster states \cite{17} is the key ingredient in a measurement-based quantum
computing (QC), i.e., one-way QC \cite{18}, with which we can carry out
expected quantum gates by some single-qubit operations and detections. The
general definition for the cluster state $\left\vert \phi_{\{k\}}\right\rangle
_{C}$ is from the set of eigenvalue equations $K^{\left(  a\right)
}\left\vert \phi_{\left\{  k\right\}  }\right\rangle _{C}=(-1)^{\kappa_{a}%
}\left\vert \phi_{\left\{  k\right\}  }\right\rangle _{C},$ with $K^{\left(
a\right)  }=\sigma_{x}^{\left(  a\right)  }\underset{b\in nghb\left(
a\right)  }{\otimes}\sigma_{z}^{\left(  b\right)  },$ and nghb(a) specifying
the sites of all the neighbors of the site $a$ and $\kappa_{a}\in\left\{
0,1\right\}  $. It has been shown that the cluster states can be created by
photons in linear optic system and by atoms going through cavities
\cite{19,20}. While for doing a one-way QC, the flying qubits, such as the
moving atoms and the photons, are not good candidates in view of an accurate
manipulation on QC. In contrast, SQUID qubits embedded in cavities are always
static, which would be good for one way QC if we could prepare them into
cluster states. \textit{W} states, whose general form is $W_{n}=(1/\sqrt
{n})\left\vert n-1,1\right\rangle $ with $\left\vert n-1,1\right\rangle $
being all the totally symmetric states involving n-1 zeros and 1 one, are
famous for their robustness to local measurement, even under qubit loss. This
makes W states useful for, e.g., quantum communication based on many nodes of
a network. A lot of schemes for \textit{W }state preparation by atoms going
through cavities have been proposed, e.g., a very recent scheme for rapidly
creating multi-atom \textit{W} states under the cavity decay \cite{21}.
Compared with the moving atoms, the SQUID qubits, without movement in the
cavities, are obviously more suitable for this job.

One of the favorable features of our schemes is that we only virtually couple
the excited level. So our implementation subspace is only spanned by the two
lowest levels of the SQUIDs, which makes our scheme simpler than previous work
\cite{10,11}, and very robust to decoherence due to spontaneous emission from
the excited level. Moreover, auxiliary qubits \cite{20} are not required in
our scheme, which improves the efficiency of our implementation. Furthermore,
we consider in our treatment some imperfect factors, such as the cavity decay
and various coupling strength of different SQUID qubits to the cavity mode and
the external microwave.

For each SQUID, two lowest flux states and an excited state are employed in
our model, as shown in Fig. 1 where quantum information is encoded in
$\left\vert 0\right\rangle $ and $\left\vert 1\right\rangle $. The spacing
between two neighboring SQUIDs is assumed to be much larger than the size of
each SQUID ring (on the order of 10-100 $\mu m$) so that the interaction
between any two SQUIDs is negligible. The SQUIDs are radiated by the cavity
field and by a classical microwave pulse which are tuned respectively to
$\omega_{c}=\left(  \omega_{02}-\delta\right)  $ and $\omega_{\mu w}=\left(
\omega_{12}-\delta\right)  $, with $\omega_{02}$ and $\omega_{12}$ resonant
frequencies between levels $\left\vert 0\right\rangle $ and $\left\vert
2\right\rangle ,$ and $\left\vert 1\right\rangle $ and $\left\vert
2\right\rangle ,$ respectively. $\delta$ is the detuning. The Hamiltonian of
the SQUID can be written in an usual form \cite{10,22}, i.e., $H_{sj}%
=\frac{Q_{j}^{2}}{2C_{j}}+\frac{(\Phi_{j}-\Phi_{xj})^{2}}{2L_{j}}-E_{J}%
\cos\left(  2\pi\frac{\Phi_{j}}{\Phi_{0}}\right)  ,$\ where the conjugate
variables $\Phi_{j}$ and $Q_{j}$ are, respectively, the magnetic flux
threading the ring and the total charge on the capacitor of the jth-SQUID,
with the commutation relation $\left[  \Phi_{j},Q_{j}\right]  =i\hbar$.
$\Phi_{xj}$ is the static external flux applied to the ring of the jth-SQUID,
and $E_{J}=I_{cj}\Phi_{0}/2\pi$ is the maximum Josephson coupling energy of
the jth-SQUID with $I_{cj}$ the critical current of the junction, and
$\Phi_{0}=h/2e$ the flux quantum. Under the action of a single-mode cavity
field $H_{c}=\omega_{c}(a^{\dagger}a+\frac{1}{2}),$\ with $\hbar=1$ assumed,
and $a^{\dagger}$ and $a$ the creation and annihilation operators of the
cavity mode, we have the interaction Hamiltonian $H_{I}$, $H_{I}=\sum
_{j=1}^{N}-\frac{1}{L_{j}}(\Phi_{j}-\Phi_{xj})(\Phi_{cj}+\Phi_{\mu wj}),$
where $\Phi_{cj}$ and $\Phi_{\mu wj}$ are the magnetic flux threading the ring
of the jth-SQUID generated by the magnetic component $\overrightarrow
{\mathbf{B}_{j}}(\overrightarrow{\mathbf{r}},t)$ of the cavity field and the
microwave pulse, respectively. $\Phi_{ij}=\int_{s_{j}}\overrightarrow
{\mathbf{B}_{j}}(\overrightarrow{\mathbf{r}},t)\cdot d\overrightarrow
{\mathbf{S}}_{j}$ with $i=c,\mu w$ corresponding to the cavity field and the
microwave pulse, respectively, and $j=1,2,\ldots,N$ regarding different qubits.

Following the steps in \cite{10}, we get to the Hamiltonian of the system in
the interaction picture $H^{I}=\sum_{j=1}^{N}g_{j}a^{\dag}e^{-i\delta
t}\left\vert 0\right\rangle _{jj}\left\langle 2\right\vert +g_{j}ae^{i\delta
t}\left\vert 2\right\rangle _{jj}\left\langle 0\right\vert +\Omega
_{j}e^{-i\delta t}\left\vert 1\right\rangle _{jj}\left\langle 2\right\vert
+\Omega_{j}e^{i\delta t}\left\vert 2\right\rangle _{jj}\left\langle
1\right\vert .$We can adiabatically eliminate the excited level $\left\vert
2\right\rangle $ by a similar way to in \cite{23} and reach an effective
Hamiltonian, $H_{I}^{\prime}=\sum_{j=1}^{N}\lambda_{j}\left(  a^{\dag
}\left\vert 0\right\rangle _{jj}\left\langle 1\right\vert +a\left\vert
1\right\rangle _{jj}\left\langle 0\right\vert \right)  ,$\ where $\lambda
_{j}=\frac{g_{j}\Omega_{j}}{\delta_{j}}$ and Max \{$g_{j},\Omega_{j}$%
\}$\ll\delta_{j}$. Considering the influence from the cavity decay, we obtain
following Hamiltonian,
\begin{equation}
H=\sum_{j=1}^{N}\lambda_{j}\left(  a^{\dag}\left\vert 0\right\rangle
_{jj}\left\langle 1\right\vert +a\left\vert 1\right\rangle _{jj}\left\langle
0\right\vert \right)  -i\frac{\kappa}{2}a\dag a,
\end{equation}
where $\kappa$ is the cavity decay rate, and we have assumed that $\kappa$ is
much smaller than $\lambda_{j}$ so that no quantum jump due to the cavity
decay actually occurs during the time evolution under our consideration. To
obtain cluster states, we assume that the N SQUIDs are prepared initially in a
product state $\left\vert \psi_{0}\right\rangle =\otimes_{j=1}^{N-1}\left\vert
+\right\rangle _{j}\left\vert 0\right\rangle _{N}$ where $\left\vert
+\right\rangle _{j}$ is the eigenstate of $\sigma_{x}^{j}$ with eigenvalue 1.
The cavity field is prepared in $\frac{1}{\sqrt{2}}\left(  \left\vert
0\right\rangle _{c}+i\left\vert 1\right\rangle _{c}\right)  ,$ which is from a
resonant interaction of an ancilla qubit initially in the state $\left(
\left\vert 0\right\rangle -\left\vert 1\right\rangle \right)  /\sqrt{2}$ with
a vacuum cavity. Before we get started, we assume the SQUIDs to be decoupled
from the cavity field, and the microwave to remain off. By adjusting the level
structure of the SQUID 1 and turning on the classical microwave to meet the
condition in Fig. 1, we reach Eq. (1) with N=1. A straightforward solution of
Eq. (1) could yield $\left\vert 0\right\rangle _{c}\left\vert 0\right\rangle
_{1}\rightarrow\left\vert 0\right\rangle _{c}\left\vert 0\right\rangle _{1},$
$\left\vert 0\right\rangle _{c}\left\vert 1\right\rangle _{1}\rightarrow
e^{-\frac{\kappa t}{4}}\{[\frac{\kappa}{4G_{1}}\sin\left(  G_{1}t\right)  +$
$\cos\left(  G_{1}t\right)  ]\left\vert 0\right\rangle _{c}\left\vert
1\right\rangle _{1}-i\frac{\lambda_{1}}{G_{1}}\sin\left(  G_{1}t\right)
\left\vert 1\right\rangle _{c}\left\vert 0\right\rangle _{1}\},$ $\left\vert
1\right\rangle _{c}\left\vert 0\right\rangle _{1}\rightarrow e^{-\frac{\kappa
t}{4}}\{[\frac{\kappa}{4G_{1}}\sin\left(  G_{1}t\right)  +\cos\left(
G_{1}t\right)  ]\left\vert 1\right\rangle _{c}\left\vert 0\right\rangle _{1}$
$-i\frac{\lambda_{1}}{G_{1}}\sin\left(  G_{1}t\right)  \left\vert
0\right\rangle _{c}\left\vert 1\right\rangle _{1}\},$ $\left\vert
1\right\rangle _{c}\left\vert 1\right\rangle _{1}\rightarrow e^{-\frac{\kappa
t}{2}}\left\vert 1\right\rangle _{c}\left\vert 1\right\rangle _{1},$ where
$G_{1}=\sqrt{\lambda_{1}^{2}-\frac{\kappa^{2}}{16}}.$\ If we choose the
evolution time of the system to be $t_{1}=[\arctan(-\frac{4G_{1}}{\kappa}%
)+\pi]/G_{1}$, the state evolution of the system is given by $\left\vert
\psi\right\rangle _{1}=\frac{1}{2}[\left\vert 0\right\rangle _{c}\left(
\left\vert 0\right\rangle _{1}+e^{-\frac{\kappa t_{1}}{4}}\frac{\lambda_{1}%
}{G_{1}}\sin\left(  G_{1}t_{1}\right)  \left\vert 1\right\rangle _{1}\right)
+i\left\vert 1\right\rangle _{c}\sigma_{z}^{1}\left(  e^{-\frac{\kappa t_{1}%
}{4}}\frac{\lambda_{1}}{G_{1}}\sin\left(  G_{1}t_{1}\right)  \left\vert
0\right\rangle _{1}+e^{-\frac{\kappa t_{1}}{2}}\left\vert 1\right\rangle
_{1}\right)  ]\otimes%
{\displaystyle\prod\limits_{j=2}^{N-1}}
\left\vert +\right\rangle _{j}\left\vert 0\right\rangle _{N}$, where
$\sigma_{z}^{1}=\left\vert 1\right\rangle _{11}\left\langle 1\right\vert
-\left\vert 0\right\rangle _{11}\left\langle 0\right\vert ,$ and the state is
not normalized if $\kappa$ is not zero. The cluster states prepared below will
be written also following this convention.

Our next step is to adjust the level structure of the SQUID 1 back to the
previous situation with decoupling to the radiation field, but adjusting the
level structure of the SQUID 2 to do the same job as done on the SQUID 1. So
we have $\left\vert \psi\right\rangle _{2}=\frac{1}{2\sqrt{2}}\{\left\vert
0\right\rangle _{c}[\left\vert 0\right\rangle _{2}\left\vert x_{1}%
\right\rangle +\sigma_{z}^{1}e^{-\frac{\kappa t_{2}}{4}}\frac{\lambda_{2}%
}{G_{2}}\sin\left(  G_{2}t_{2}\right)  \left\vert 1\right\rangle
_{2}\left\vert y_{1}\right\rangle ]+$ $i\left\vert 1\right\rangle _{c}%
\sigma_{z}^{2}[e^{-\frac{\kappa t_{2}}{4}}\frac{\lambda_{2}}{G_{2}}\sin\left(
G_{2}t_{2}\right)  \left\vert 0\right\rangle _{2}\left\vert x_{1}\right\rangle
+\sigma_{z}^{1}e^{-\frac{\kappa t_{2}}{2}}\left\vert 1\right\rangle
_{2}\left\vert y_{1}\right\rangle ]\}\otimes%
{\displaystyle\prod\limits_{j=3}^{N-1}}
\left\vert +\right\rangle _{j}\left\vert 0\right\rangle _{N}$, where
$\left\vert x_{1}\right\rangle =\left\vert 0\right\rangle _{1}+$
$e^{-\frac{\kappa t_{1}}{4}}\frac{\lambda_{1}}{G_{1}}\sin\left(  G_{1}%
t_{1}\right)  \left\vert 1\right\rangle _{1},$ $\left\vert y_{1}\right\rangle
=e^{-\frac{\kappa t_{1}}{4}}\frac{\lambda_{1}}{G_{1}}\sin\left(  G_{1}%
t_{1}\right)  \left\vert 0\right\rangle _{1}+e^{-\frac{\kappa t_{1}}{2}%
}\left\vert 1\right\rangle _{1}$\ and $G_{2}=\sqrt{\lambda_{2}^{2}%
-\frac{\kappa^{2}}{16}}$ is the effective coupling regarding the SQUID 2,
including the cavity decay.\ Step by step we may carry out above operation on
the rest SQUIDs individually. After the (N-1)-th SQUID is performed by the
same operation as on the first one, we have%
\begin{equation}
\left\vert \psi\right\rangle _{N-1}=\frac{1}{2^{N/2}}(\left\vert
0\right\rangle _{c}\left\vert x_{N-1}\right\rangle +i\left\vert 1\right\rangle
_{c}\sigma_{z}^{N-1}\left\vert y_{N-1}\right\rangle )\otimes\left\vert
0\right\rangle _{N}\text{,}%
\end{equation}
where $\left\vert x_{N-1}\right\rangle $ and $\left\vert y_{N-1}\right\rangle
$ are two different entangled states of the first (N-1) SQUIDs, and they
follow recursive relations below $\left\vert x_{N-1}\right\rangle =\left\vert
0\right\rangle _{N-1}\left\vert x_{N-2}\right\rangle +\sigma_{z}%
^{N-2}e^{-\frac{\kappa t_{N-1}}{4}}\frac{\lambda_{N-1}}{G_{N-1}}\sin\left(
G_{N-1}t_{N-1}\right)  \left\vert 1\right\rangle _{N-1}\left\vert
y_{N-2}\right\rangle ,$ $\left\vert y_{N-1}\right\rangle =e^{-\frac{\kappa
t_{N-1}}{4}}\frac{\lambda_{N-1}}{G_{N-1}}\sin\left(  G_{N-1}t_{N-1}\right)
\left\vert 0\right\rangle _{N-1}\left\vert x_{N-2}\right\rangle +\sigma
_{z}^{N-2}e^{-\frac{\kappa t_{N-1}}{2}}\left\vert 1\right\rangle
_{N-1}\left\vert y_{N-2}\right\rangle .$ For the Nth SQUID we stop the state
evolution at $t_{N}=\frac{1}{G_{N}}[\arctan(-\frac{4G_{N}}{\kappa})+\pi]$,
with $G_{N}=\sqrt{\lambda_{N}^{2}-\frac{\kappa^{2}}{16}},$ then can obtain%

\begin{equation}
\left\vert \psi\right\rangle _{N}=\frac{1}{2^{N/2}}\left(  \left\vert
0\right\rangle _{N}\left\vert x_{N-1}\right\rangle +\frac{\lambda_{N}}{G_{N}%
}e^{-\frac{\kappa t_{N}}{4}}\sin\left(  G_{N}t_{N}\right)  \left\vert
1\right\rangle _{N}\sigma_{z}^{N-1}\left\vert y_{N-1}\right\rangle \right)
\otimes\left\vert 0\right\rangle _{c}\text{.}%
\end{equation}
To check above state, we first assume $\kappa=0,$ i.e., the ideal condition.
In this case, Eq. (3) reduces to the standard cluster state \cite{17}.
\begin{equation}
\left\vert \Phi\right\rangle _{N}=\frac{1}{2^{N/2}}\otimes_{j=1}^{N}\left(
\left\vert 0\right\rangle _{j}+\left\vert 1\right\rangle _{j}\sigma_{z}%
^{j-1}\right)  .
\end{equation}

To evaluate our prepared cluster state under the cavity decay, we have
calculated the fidelity F and the success probability P, as shown in Fig. 2
where we considered the qubit numbers 2, 3, and 4, respectively.\ It is
evident that both F and P are going down with the qubit number and $\kappa.$
It is understandable that the preparation of a cluster state with bigger size
need longer time and is thereby affected more by the cavity decay. Thus in the
presence of the cavity decay, the size of the cluster states are limited. For
example, if we want $F>0.95,$ in the case of $\lambda_{1}=\lambda
_{2}=...=\lambda,$the maximal qubit number of the cluster state is 32,
corresponding to $F=0.951$. For a cluster state with more qubit, we could
created it step by step by connecting the few-qubit cluster states by quantum
phase gates on the end qubits of the neighboring SQUID chains \cite{24}.
Moreover, to keep the system as stable as possible, we should not change the
level spacing of each SQUID drastically. In this sense, we estimate the
appearance of imperfection due to the time delay of the level spacing
adjustment, i.e., the undesired phases regarding the component states of the
superposition. This estimate is plotted also in Fig. 2.

From now on, we turn to the generation of \textit{W} states using SQUIDs. In
contrast to the cluster state preparation by individually manipulating the
SQUIDS, we have to simultaneously couple all the SQUIDs in this case to the
cavity mode and to the microwave. Provided that the SQUIDs and the cavity mode
are initially in the state $%
{\textstyle\prod\limits_{j=2}^{N}}
\left\vert 1\right\rangle _{1}\left\vert 0\right\rangle _{j}$ and the vacuum
state $\left\vert 0\right\rangle _{c}$, respectively, we have the time
evolution of the system by straightforwardly solving Eq. (1),%
\begin{align}
\left\vert \psi(t)\right\rangle  &  =%
{\textstyle\prod\limits_{j=2}^{N}}
\left\vert 1\right\rangle _{1}\left\vert 0\right\rangle _{j}\left\vert
0\right\rangle _{c}\{1+\frac{\lambda_{1}^{2}}{A^{2}}[-1+e^{-\frac{\kappa t}%
{4}}(\cos\frac{Bt}{4}+\frac{\kappa}{B}\sin\frac{Bt}{4})]\}\nonumber\\
&  +\sum_{k=2}^{N}\left\vert 1_{k}\right\rangle
{\textstyle\prod\limits_{j=1,j\neq k}^{N}}
\left\vert 0\right\rangle _{j}\left\vert 0\right\rangle _{c}\frac{\lambda
_{1}\lambda_{k}}{A^{2}}[-1+e^{-\frac{\kappa t}{4}}(\cos\frac{Bt}{4}%
+\frac{\kappa}{B}\sin\frac{Bt}{4})]\nonumber\\
&  -%
{\textstyle\prod\limits_{j=1}^{N}}
\left\vert 0\right\rangle _{j}\left\vert 1\right\rangle _{c}\frac
{i4\lambda_{1}}{B}\sin\frac{Bt}{4}e^{-\frac{\kappa t}{4}}\text{.}%
\end{align}
where $A^{2}=%
{\textstyle\sum\limits_{j=1}^{N}}
\lambda_{j}^{2}$, and $B=\sqrt{16A^{2}-\kappa^{2}}$. In order to obtain
\textit{W} states of (N-1) SQUIDs, we have the conditions $t=\frac{4\pi}{B}$,
and $\lambda_{1}^{2}=A^{\prime2}e^{\frac{\kappa t}{4}}$ to be satisfied. Then
the system evolves to%

\begin{equation}
W_{N-1}=e^{-\frac{\kappa t}{8}}\sum_{k=2}^{N}\frac{\lambda_{k}}{A^{\prime}%
}\left\vert 1_{k}\right\rangle
{\textstyle\prod\limits_{j=2,j\neq k}^{N}}
\left\vert 0\right\rangle _{j},
\end{equation}
where $A^{\prime2}=%
{\textstyle\sum\limits_{k=2}^{N}}
\lambda_{k}^{2}$. This is a \textit{W} state of (N-1) SQUIDs with arbitrary
coefficients. By choosing the same coupling for the rest (N-1) SQUIDs except
the first one, we may produce a standard \textit{W} state%

\begin{equation}
W_{N-1}=e^{-\frac{\kappa t}{8}}\frac{1}{\sqrt{N-1}}\sum_{k=2}^{N}\left\vert
1_{k}\right\rangle
{\textstyle\prod\limits_{j=2,j\neq k}^{N}}
\left\vert 0\right\rangle _{j},
\end{equation}
with the success probability $P=e^{-\frac{\kappa t}{4}}$. We plot in Fig. 3
the success probability $P$ versus cavity decay $\kappa$. From above condition
for the evolving time, we have $\lambda_{1}=A^{\prime}e^{\kappa\pi/2B}$ $\geq$
$A^{\prime},$ which means that for many-qubit case, the coupling strength
regarding the first qubit is much larger than those of others. This also
implies that, the coupling strength regarding other SQUIDs would be quite
small if $\lambda_{1}$ is not big. Experimentally, the difference of
$\lambda_{1}$ from other coupling strength could be made by reducing the
detuning regarding the first SQUID. Alternatively, we may put the first SQUID
on the antinode of the cavity standing wave (i.e., the maximum coupling), but
others on the positions deviated from corresponding antinodes.

We address below the experimental feasibility of our schemes. The
implementation time of our scheme should be much shorter than the cavity decay
time $\kappa^{-1}=Q/2\pi\nu_{c},$ where Q is the quality factor of the cavity
(Q= $10^{6}-10^{8}$ has been achieved experimentally \cite{25}, and for
superconducting qubits \cite{26}), and $\nu_{c}$ is the cavity field
frequency. The coupling constants of the SQUID to the cavity field and to the
classical microwave available at present are $g\sim1.8\times10^{8}s^{-1}$, and
$\Omega\sim8.5\times10^{7}s^{-1}$, respectively \cite{27}. So we may have an
effective coupling $\lambda\sim10$ MHz if assuming $\delta\sim1.5$ GHz$.$ In
this case, the interaction time $\tau$ is on the order of $10^{-7}\sec$, much
shorter than $\kappa^{-1}\sim4\times10^{-5}$ (with $Q=10^{7}$ and $\nu
_{c}=40GHz$). This implies that our proposed states can be of the fidelity
higher than 0.99 due to $\kappa/\lambda=4\times10^{-2}.$

Although optical photons in cluster states have been experimentally
demonstrated some important features for one-way QC, we prefer fixed qubits,
like the SQUIDs embedded in the cavity, for storing and processing quantum
information. This statement is also applicable to the \textit{W} state case
for quantum information processing. Moreover, due to no necessity of auxiliary
qubits, our scheme could much reduce the experimental challenge for cluster
state preparation. For example, generation of a two-qubit cluster state
requires at least five steps in \cite{19} due to the auxiliary states
involved, while only two steps are needed in our scheme. Furthermore, we only
virtually couple the excited levels throughout our scheme, which, compared to
\cite{10,11}, could both simplify the implementation steps and reduce the
possibility of decoherence due to spontaneous emission. We have given the
analytical expressions of the prepared cluster state and W state under the
cavity decay, and numerically assessed the related fidelities.

Before ending our discussion, we give few comments on our wavefunction
treatment for the cavity decay. The quantum-jump approach \cite{28} has been
widely employed in solving dissipative dynamics for quantum systems, which
lies in the time evolution governed by a non-Hermitian operator and
interrupted by instantaneous jumps by the detection of a photon. Normally, a
solution by quantum-jump approach has to be resorted to numerical
calculations. In our case, however, to prepare high-fidelity cluster states
and \textit{W} states, we prefer our implementation time much shorter than the
cavity decay time. Therefore, we may only focus on the time evolution of the
system governed by the non-Hermitian Hamiltonian Eq. (1)\ before the leakage
of photons occurs. In this sense, our wavefunction method would be
advantageous over the quantum-jump approach in the availability of analytical
solution clearly demonstrating the detrimental influence from the cavity decay
on the prepared states. So our solutions might be helpful in experiments for
estimating the infidelity and correcting the error, particularly for
implementation with cavities of comparatively low quality. This also implies
that our numerical results in the last two figures are valid only within the
regime of $\kappa/\lambda\ll1$ $($\textit{e.g.,} $\kappa/\lambda
\leqslant0.1).$

In summary, we have proposed potential schemes for creating cluster states and
\textit{W states} of many SQUIDs. Fast adjustments of the level spacings of
individual SQUIDs are needed in the generation of the cluster states. As no
auxiliary qubits or flying qubits involved, our scheme gives good candidates
for one-way QC. In the generation of the \textit{W states, all} the SQUIDs are
coupled to the radiation fields simultaneously, which results in that the
cavity decay only affects the prepared states globally, instead of on the
internal structure of the W state. Throughout our schemes, the excited level
is only virtually coupled, and we have specifically studied the detrimental
influence from the cavity decay. Our analytical results have shown clearly
that the cluster states and the \textit{W} states may be generated with
high-fidelity only in the case of tiny cavity decay rate.

This work is supported by National Natural Science Foundation of China under
Grants No. 10474118 and No. 60490280, by Hubei Provincial Funding for
Distinguished Young Scholars, and by the National Fundamental Research Program
of China under Grants No. 2005CB724502.

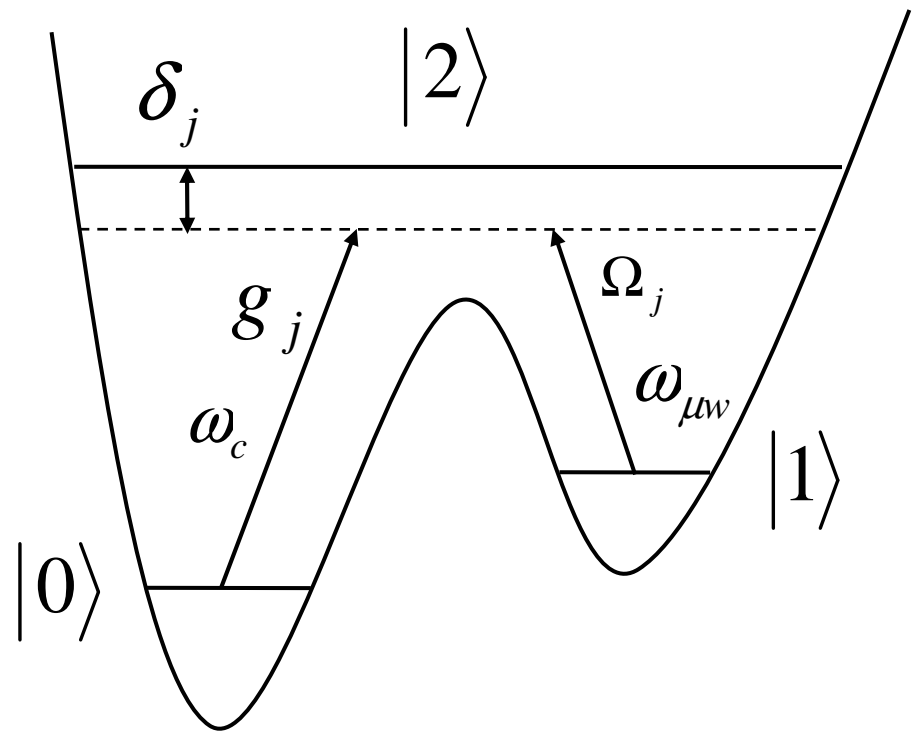
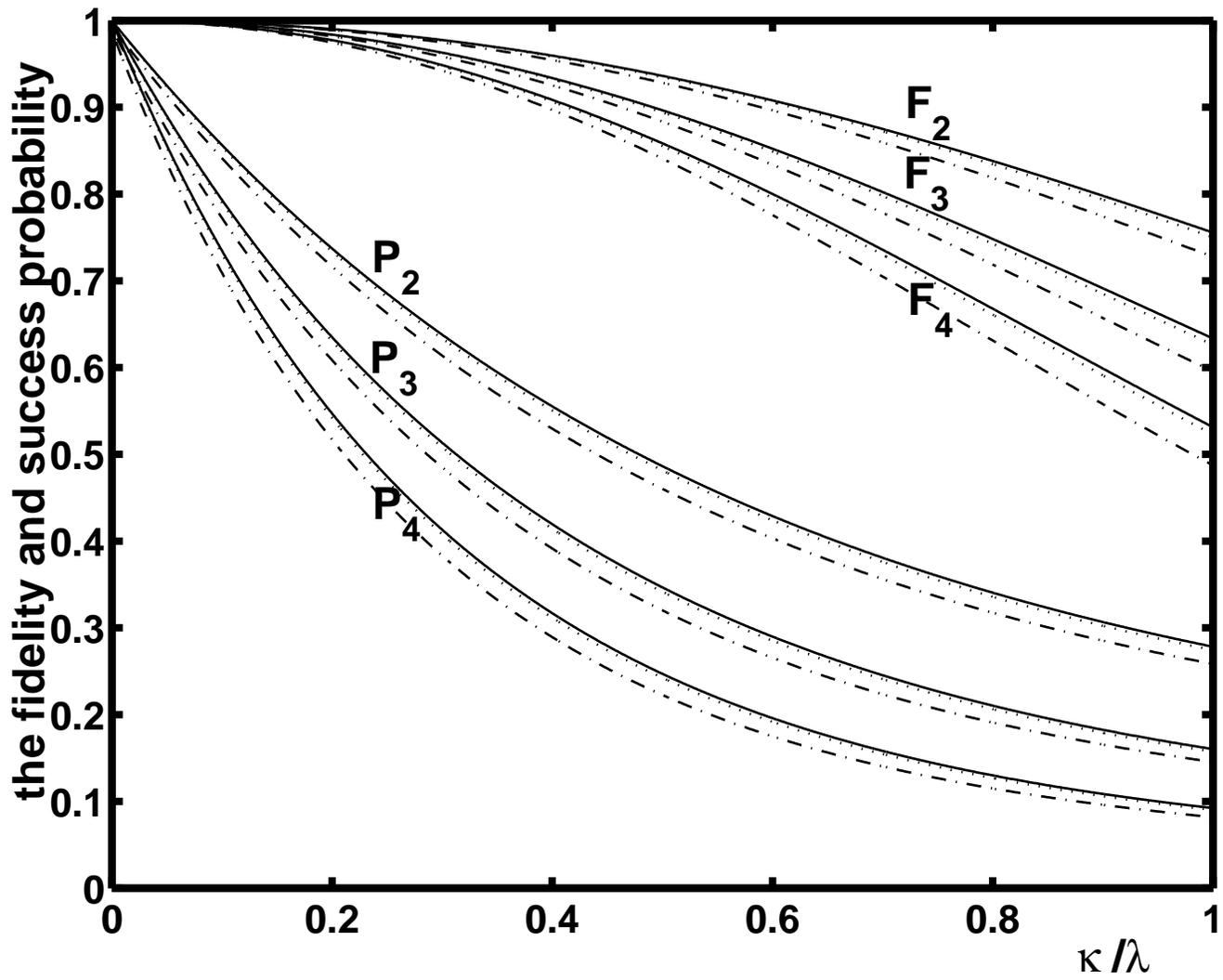
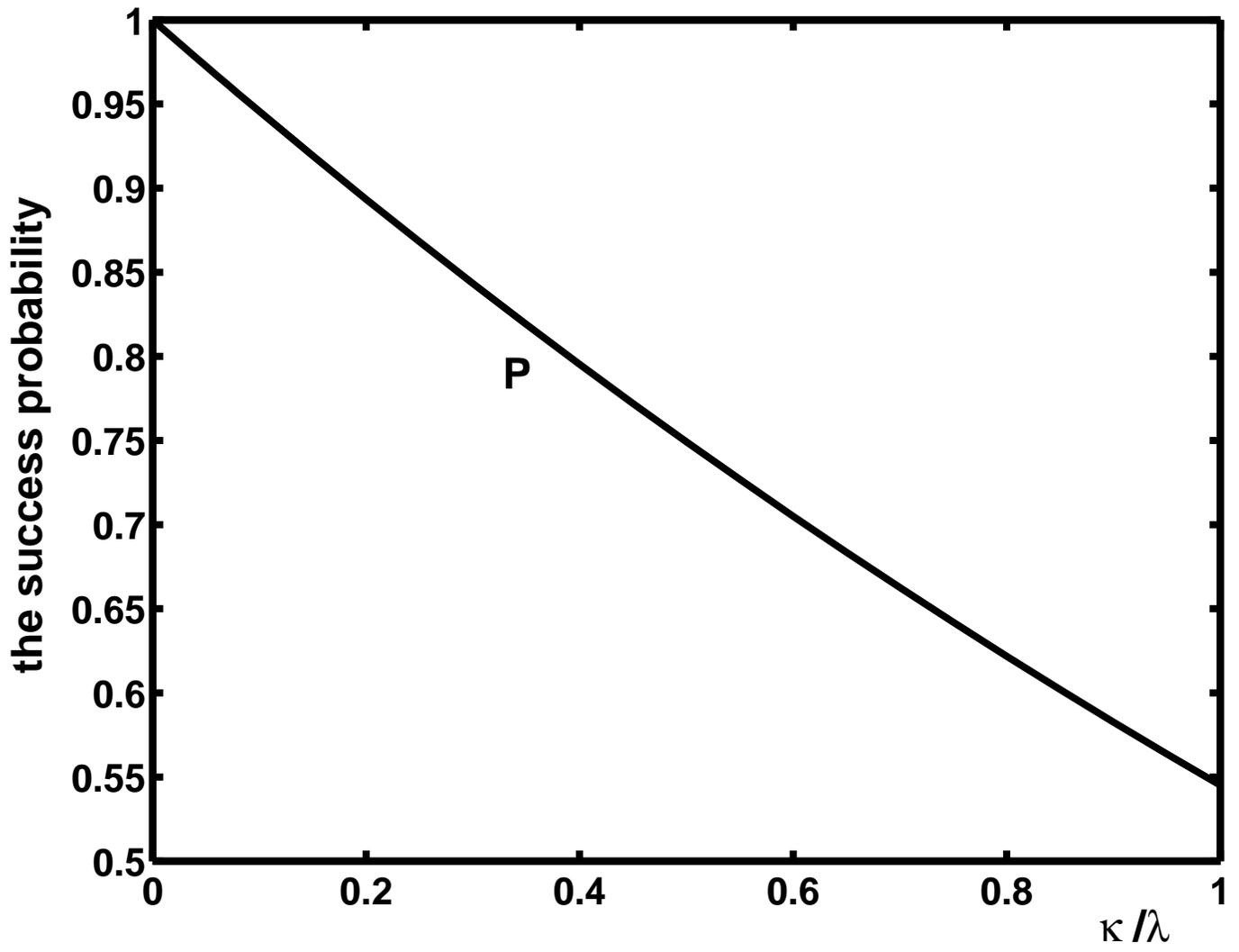
\end{document}